\documentclass[preprintnumbers,amsmath,amssymb,twocolumn,pra]{revtex4}

\usepackage{graphicx,comment}
\usepackage{epstopdf}
\usepackage{dcolumn}
\usepackage{bm}
\usepackage[T1]{fontenc}
\usepackage[latin1]{inputenc}
\usepackage{enumerate}
\usepackage{fancyhdr,color}
\usepackage{verbatim}

\input{Qcircuit}

\def\cH{\mathcal{H}}

\def\cO{\mathcal{O}}

\def\cU{\mathcal{U}}

\def\Tr{\mathrm{Tr}}
\def\Pr{\mathrm{Pr}}

\def\one{{\mathchoice {\rm 1\mskip-4mu l} {\rm 1\mskip-4mu l} {\rm
1\mskip-4.5mu l} {\rm 1\mskip-5mu l}}}
\newcommand{\ketbra}[1]{| #1 \rangle \! \mspace{2mu}\langle #1| }

\begin{document}

LA-UR-13-23134

SAND  2013-5836J
\title{On the  gap of Hamiltonians for the adiabatic simulation of quantum circuits}

\author{Anand Ganti}
\email{aganti@sandia.gov}

\affiliation{
Sandia National Laboratories \\
Albuquerque, New Mexico 87185, USA \\
}
\author{Rolando D. Somma}
\email{somma@lanl.gov}

\affiliation{
Los Alamos National Laboratory \\
Los Alamos, New Mexico 87545, USA 
}

\begin{abstract}
The time or cost of simulating a quantum circuit by adiabatic evolution is determined by  
the spectral gap of the Hamiltonians involved in the simulation. In ``standard'' constructions
based on Feynman's Hamiltonian,
such a gap decreases polynomially with the number of gates in the circuit, $L$.
Because a larger gap
implies a smaller cost, we study  the limits of spectral gap amplification in this context.
We show that, under some  assumptions on the ground states and the cost of evolving with the Hamiltonians
 (which apply to the standard constructions), an upper bound
on the gap of order $1/L$ follows.
In addition, if the Hamiltonians satisfy a frustration-free property, 
 the upper bound is of order $1/L^2$. Our proofs use recent results on adiabatic state transformations, spectral gap amplification, and the simulation of continuous-time quantum query algorithms. They also consider a reduction from the unstructured search problem, whose lower bound in the 
 oracle cost translates into the upper bounds in the gaps. The impact of our results is that improving the gap beyond that of standard constructions (i.e., $1/L^2$), if possible, is challenging.
 
%
%
%
%
 
%
%
%

\end{abstract}

\date{\today}
             
\maketitle

\section{Introduction} 
\label{sec:intro}
Adiabatic quantum computing (AQC) is an alternative to the standard circuit 
model of quantum computation. In AQC, the input  is a (qubit) Hamiltonian $H(1)$ and
the goal is to prepare the ground state  of $H(1)$
by means of slow or adiabatic evolutions. One then sets an initial Hamiltonian $H(0)$
and  builds a Hamiltonian path $H(g)$, $0 \le g \le 1$, that 
interpolates between $H(0)$ and $H(1)$. If the ground states
of $H(g)$ are continuously related and
remain at a spectral gap of order $\Delta$ with any other eigenstate during the evolution, 
the quantum adiabatic
approximation implies that, for 
\begin{align}
\label{eq:runningtime}
\dot g(t) \le   \epsilon \frac{\Delta^q}{h} \; ,
\end{align}
the ground state of $H(1)$ can be adiabatically prepared with fidelity $1-\epsilon$.
$0<q \le 3$ and $h$ depends on $\| \partial^n H(g)/\partial g^n \|^{q+1}$, $n=1,2$, for differentiable paths
\cite{messiah_1999,jansen_bounds_2007,regev_quantum_2004,lidar_adiabatic_2009,boixo:qc2009b}.

A key feature of AQC is that it constitutes a ``natural'' model for  problems that efficiently reduce
to the computation of ground-state properties. Some of these are problems
in combinatorial optimization \cite{finnila:qc1994a,kadowaki:qc1998a,farhi_quantum_2000,farhi:qc2001a,
somma_thermod_2007,somma_optimization_2010,somma_quantum_2008} and problems
in many-body physics, e.g. the computation of a quantum phase
diagram~\cite{sachdev_2001}. Whether AQC is robust to decoherence
or not is unclear and a complete fault-tolerant  implementation of AQC remains unknown~\cite{jordan_adiabatic_2006,lidar_FTAQC_2008}.  Nevertheless, the role of the spectral gap
is  imperative in  a noisy implementation of AQC:  a bigger $\Delta$ could imply
a smaller running time [Eq.~\eqref{eq:runningtime}] and a reduction of the 
(unwanted) population of excited states due to thermal effects. 
Our goal is then to study the limits and possibilities of amplifying the gap in AQC.
Roughly stated, we are  addressing  the following question: Given $H(g)$ with gap $\Delta(g)$ and ground state 
$\ket{\psi(g)}$, can we find $\tilde H(g)$ with gap $\tilde \Delta(g) \gg \Delta(g)$ and ground state
satisfying $|\tilde \psi(g)\rangle \approx \ket{\psi(g)}$? 
Our motivation is the same as that of
Ref.~\cite{mizel_fixedgap_2010}.
We are particularly interested in amplifying the gap of those 
Hamiltonians that arise in the adiabatic simulation of general quantum circuits-- see below.

The power of AQC and the standard quantum circuit model
are equivalent. That is, any  algorithm in the AQC model with
a running time $T$, that prepares a quantum state
$\ket {\psi(1)}$, can be simulated with a quantum circuit
of $L \in {\rm poly}(T)$ unitary gates that prepares
a sufficiently close state to ${\ket {\psi(1)}}$ when acting on some trivial initial state~\cite{berry_efficient_2007,
wiebe_product_2010,cleve_query_2009,childs_efficient_2010}. The converse also holds:
Any quantum circuit of $L$ unitary gates that prepares a quantum state
$\ket {\phi^L}$, when acting on some trivial initial state, 
can be simulated within the AQC model by evolving adiabatically with 
suitable Hamiltonians $H(g)$ for time $T \in {\rm poly}{ (L)}$. 
The ground state of the final Hamiltonian, $\ket{\psi(1)}$, has a large probability
of being in $\ket {\phi^L}$ after a simple measurement~\cite{aharonov_adiabatic_2007,
mizel_equivalence_2007,aharonov_line_2009}. $H(g)$ depends on the unitaries
that specify the quantum circuit.

To describe our results in detail, 
we  review the first ``standard'' construction in Ref.~\cite{aharonov_adiabatic_2007},
which is based on Feynman's Hamiltonian~\cite{feynman_simulating_1982}.
 $\cU$ is  a quantum
circuit acting on $n$ qubits that prepares the ``system''  state $\ket {\phi^L}= U^L \ldots U^1 \ket {\phi^0}$,
after the action of $L$ unitary gates $U^1,\ldots,U^L$. 
There is also an ancillary system, denoted by ``clock'',
whose
basis states are $\{ \ket 0_{\rm c} , \ket 1_{\rm c}, \ldots , \ket L_{\rm c} \}$. The (final) Hamiltonian $H^\cU$
is mainly a sum of two terms.
The first term is the so-called Feynman Hamiltonian:
\begin{align}
\nonumber
 &H_{\rm Feynman}^\cU =  \sum_{l=1}^L  h^{\cU,l} \; , \\
 \label{eq:Hfeynman}
& h^{\cU,l} = \frac 1 2 ( \one \otimes  \ketbra l_{\rm c} +\one \otimes \ketbra{l-1}_{\rm c} -  \\
\nonumber
&  - U^l \otimes \ket l \! \bra{l-1}_{\rm c} - (U^l)^\dagger \otimes \ket {l-1}\! \bra{l} _{\rm c} )  \; .
\end{align}
$\one$ is the trivial operation on the system's state. $H_{\rm Feynman}^\cU$ 
can be easily diagonalized by using the Fourier transform.
The eigenvalues are $1- \cos k$, with $k = 2 \pi m/(L+1)$ and $m \in \mathbb{Z}$. Then, the 
lowest eigenvalue is zero ($k=0$) and the gap is  of order $1/L^2$ [$k=2\pi/(L+1)$
for the smallest nonzero eigenvalue].
Each eigenvalue appears with multiplicity $2^n$, corresponding to each state $\ket \sigma$ of the system.  
The eigenstates of $H_{\rm Feynman}^\cU$ are
\begin{align}
\label{eq:feynmaneigenstate}
\frac 1 {\sqrt{L+1}}\sum_{l=0}^L e^{i kl }U^l \ldots U^0 \ket{\sigma} \otimes \ket l_{\rm c} \; ,
\end{align}
 where $U^0 = \one$.  
 If $\ket \sigma= \ket {\phi^0}$,
 the  eigenstate in Eq.~\eqref{eq:feynmaneigenstate} 
 has probability  $1/(L+1)$ of being in the state  output by the circuit.
 That is, we can prepare $\ket {\phi^L}$ with such a probability
 by a projective measurement of the clock register on the state of Eq.~\eqref{eq:feynmaneigenstate}. 
 We can remove the multiplicity of the lowest eigenvalue
 if we add a second term, $ H_{\rm input}$,  whose expected value vanishes when 
 $\ket \sigma = \ket {\phi^0}$ and is strictly positive otherwise.
For example, if $\ket {\phi^0} = \ket+^{\otimes n}$, where $\ket + = (\ket 0 + \ket 1)/\sqrt 2$, 
 $H_{\rm input}$ in Ref.~\cite{aharonov_adiabatic_2007} corresponds to
 \begin{align}
 \nonumber
 H_{\rm input} = \sum_{j=1}^n \ketbra{-} _j\otimes \ketbra 0 _{\rm c} \; ,
 \end{align}
with $\ket - = (\ket 0 - \ket 1)/\sqrt 2$.
In this case, $ H_{\rm input}$   sets a ``penalty''  if the system-clock initial state 
 is different from $\ket + ^{\otimes n} \otimes \ket 0 _{\rm c}$.  
 The lowest eigenvalue of $H_{\rm input} $ is zero and the gap is a constant independent of $L$ (i.e., $\Delta_{\rm input}=1$).

Then, the Hamiltonian
 \begin{align}
 \label{eq:standardH}
 H^\cU= H_{\rm Feynman}^\cU + H_{\rm input} 
 \end{align}
  has 
\begin{align}
\label{eq:historystate}
\ket {\psi^\cU} = \frac 1 {\sqrt{L+1}}\sum_{l=0}^L\ket {\phi^l}\otimes \ket l_{\rm c}
\end{align}
as unique ground state [$k=0$ in Eq.~\eqref{eq:feynmaneigenstate}], where $\ket{\phi^l}=U^l \ldots U^0 \ket{\phi^0}$. 
We will refer to  $\ket {\psi^\cU}$   as the ``history state''.
The lowest eigenvalue of $H^\cU$ is also zero and the spectral gap satisfies $\Delta^\cU \in \Theta(1/L^2)$
\cite{deift_improved_2007}. It is simple to construct an interpolating path $H^\cU(g)$
that has a spectral gap $\Delta^\cU(g) =\Delta^\cU \in \Theta(1/{\rm poly}L)$ for all $g$ and $H^\cU(1)=H^\cU$.
This is done by, for example, parametrizing the unitaries in the
circuit so that $U^l \rightarrow U^l(g)$ in Eq.~\eqref{eq:standardH}, and $U^l(0)=\one$, $U^l(1)=U^l$.
Then, the ground state  $\ket {\psi^\cU(1)}=\ket{\psi^\cU}$ can be prepared from $\ket {\psi^\cU(0)} = \ket{\phi^0} \otimes \sum_l \ket l_{\rm c}/\sqrt{L+1}$
by evolving adiabatically with $H^\cU(g)$ for time $T \in \cO[{\rm poly}(L)] $ [see Eq.~\eqref{eq:runningtime}].

$H^\cU$ is  often regarded as ``unphysical'' as the system-clock interactions
may represent non-local interactions of actual quantum subsystems (qubits).
Then, a number of steps that include modifications of the gates in the circuit
and techniques from perturbation theory (e.g., perturbation gadgets)~\cite{nagaj_2008},  allow us to reduce $H^\cU$ 
to a physical, local Hamiltonian $H^\cU_{\rm local}$.
Such steps preserve
  the two main ingredients for showing the equivalence between AQC and
  the circuit model: i- that the spectral gap of the local Hamiltonian, $\Delta^\cU_{\rm local}$, is  bounded from below by
   $1/{\rm poly}(L)$ and ii- that the ground state  has sufficiently large probability of being in 
   $\ket{ \phi^L}$  after a simple quantum operation (e.g., a simple projective measurement). It is important to remark that  $\Delta^\cU_{\rm local}$ is smaller than $\Delta^\cU$
   in  standard constructions~\cite{aharonov_adiabatic_2007,aharonov_line_2009}. For this reason, some
   attempts to improve the running time of the adiabatic simulation of a quantum circuit
   consider first the amplification of $\Delta^\cU$
   by making simple modifications to  $H^\cU$ (see Ref.~\cite{lloyd_adiabatic_2008} for an example);
   Our results concern the amplification of $\Delta^\cU$.
   
   In this report we show that, under some assumptions on the ground states and the time or
   cost of evolving with the Hamiltonians, an upper bound on the gap of order $1/L$ follows.
   Furthermore, if the Hamiltonians additionally satisfy a so-called frustration-free property, then the upper 
   bound is $1/L^2$. An implication of our results is that  simple modifications to  $H^\cU$ in Eq.~\eqref{eq:standardH}
   are not sufficient to amplify its gap. 
   While such modifications could be useful to prepare the desired state via a constant-Hamiltonian evolution~\cite{landahl_PST_2004},
   they may not be useful to prepare the state adiabatically.  
   Our proofs are constructive, i.e., we find a reduction from the unstructured search problem~\cite{grover_fast_1996}
   (Sec.~\ref{sec:grover}),
   whose lower bound on the oracle cost~\cite{bennett_searchbound_1997} (i.e.,
   the number of queries to the oracle needed) can be transformed into the upper bounds on the gaps
   (Sec.~\ref{sec:mainresult}). 
   Clearly, the only way to obtain a bigger gap, if gap amplification is indeed possible, 
   is by avoiding one or more assumptions needed for our proofs. 
   This suggests a migration from those constructions that are based on Feynman's Hamiltonian.
       
  \section{Search by a generalized measurement-based method}
  \label{sec:grover}
  The proof of an upper bound on $\Delta^\cU$ uses a reduction from the unstructured
 search problem or SEARCH. In this section, we show a quantum method
 that solves SEARCH using measurements.

 For a system of $n$ qubits,
 we let $N=2^n$ be the dimension of the associated state (Hilbert) space  $\cH$. 
 Given an oracle $O_X$, where the 
 input $X$ is a $n$-bit string, the goal of SEARCH is to output $X$.
 In quantum computing,  $O_X$ implements the following unitary operation:
 \begin{align}
 \nonumber
 O_X \ket Y= \left\{  \begin{matrix}
&  \ \ \ \ket {Y} \ {\rm if} \ X \ne Y \; , \cr &- \ket X \  {\rm if  } \ X = Y \; .
 \end{matrix} \right.
 \end{align}
 $O_X$ acts on ${\cal H}$. A quantum algorithm for SEARCH uses $O_X$ and other $X$-independent operations
 to prepare a state sufficiently close to $\ket X$. Thus, a projective measurement on 
 this state outputs $X$ with large probability. The (oracle) cost of the algorithm is given 
 by the number of times that $O_X$ is used. A lower bound $\Omega(\sqrt N)$
  for the cost of SEARCH is known~\cite{bennett_searchbound_1997} and the famous Grover's algorithm
 solves SEARCH with $L/2 \in \Theta(\sqrt N)$ oracle uses \cite{grover_fast_1996}.   
 Grover's algorithm, denoted by $\cU_X$, is a sequence of two unitary operations,
 $O_X$ and $R$, where $R$ is a reflection over the equal superposition state $\ket {\phi^0}= \frac 1 {\sqrt{N}} \sum_Y \ket Y=\ket{+}^{\otimes n}$.
 The initial state is also $\ket {\phi^0}$.
 The state output by $\cU_X$ is  $\ket {\phi^L_X}$ and satisfies, in the large $N$ limit,  
 \begin{align}
 \label{eq:halfgroverstate}
\left \| \ket {\phi_X^L} -  \ket X ] \right \| \ll 1 \; .
 \end{align}
 
There are other quantum methods that solve SEARCH,
 with optimal cost, using measurements. 
 One such method, first introduced in Ref.~\cite{childs_quantum_2002}, involves
 two projective measurements: After preparing $\ket{\phi^0}$, a  measurement
 of $\ket{\psi_X} \approx[\ket X + \ket{\phi^0}]/\sqrt 2$, followed by a measurement of $\ket{X}$, outputs $X$ with probability
 close to $1/4$. The cost of this measurement-based method is dominated
 by the simulation of the first measurement. 
 Such a  simulation can be done using the phase estimation algorithm~\cite{kitaev_quantum_1995}
 or by phase randomization~\cite{boixo:qc2009a}. Both methods require evolving
 with a Hamiltonian that has $\ket {\psi_X}$ as eigenstate.
 The  evolution time is proportional to the inverse  gap of the Hamiltonian, which is needed to resolve
 the desired state from any other eigenstate.
 
 Generalizations of the above measurement-based method, that consider  simulating measurements
  in other states, also solve SEARCH. To see this, we let $\ket{\zeta_X} \in \cH'$ and $\ket \nu \in \cH''$   
  be two pure quantum states that do and do not depend on $X$, respectively. 
  The corresponding Hilbert spaces satisfy $\cH'' \subseteq \cH' \subseteq \cH$.
  We also define 
   \begin{align}
 \nonumber
p_{\nu, \zeta_{X}} &= {\rm tr} [ \langle \zeta_X\ketbra{\nu} \zeta_X \rangle ]  \; , \\
\nonumber
p_{X, \zeta_{X}} &= {\rm tr} [ \langle{\phi^L_X}  \ketbra{\zeta_{X}}\phi^L_X \rangle ]   \; ,
 \end{align}
 which are the probabilities of projecting $\ket \nu$ into $\ket{\zeta_X}$, and $\ket{\zeta_X}$ into $\ket X$, respectively,
 after a measurement (on the corresponding Hilbert spaces).
A generalization of the measurement-based method is described in Table I.
The probability of success is  $p_s \ge p_{\nu, \zeta_{X}} \cdot p_{X, \zeta_{X}}$. 

\vspace{0.5cm}
\begin{tabular}{m{8cm} c}
\label{GMBM}
Table I. Generalized measurement-based method\\
\hline
\hline
\begin{description}
\item[i-   ]   Prepare $\ket \nu$
\item[ii- ] Measure $\ket{\zeta_X}$
\item[iii-] Measure $\ket X$
\end{description} \\
\hline
\end{tabular}
\vspace{0.5cm}

  We  now obtain the time $T$ of solving SEARCH (with probability $p_s$) with
  the generalized measurement-based method.
  We let  $G_{X}$ be the Hamiltonian that has $\ket {\zeta_{X}}$
 as unique ground state and the corresponding spectral gap of $G_X$ is $\Delta^{\cU_X}$.
 $G_X$ acts on the Hilbert space ${\cal H}'$ and depends on $O_X$.
  $T$ is determined by the total time of evolution with $G_X$ needed
 to simulate the measurement of $\ket{\zeta_X}$ in step ii.    Using
 the phase estimation algorithm~\cite{kitaev_quantum_1995}  or  evolution randomization~\cite{boixo:qc2009a},
 this time is
 \begin{align}
 \label{eq:GMBcost}
 T = c/\Delta^{\cU_X} \; ,
 \end{align}
 for some constant $c\ge \pi$.
The lower bound in the oracle cost of SEARCH 
can then be used to set a lower bound on $T$ or, equivalently, an upper bound
in $\Delta^{\cU_X}$. This results from noting that the evolution under $G_X$ can be well approximated
with a discrete sequence of unitaries that contains $O_X$.
Nevertheless, to make a rigorous statement on $\Delta^{\cU_X}$, some assumptions on $G_X$
and the ground state are needed. We provide such assumptions and  our main results in the next section.

  


  \section{Gap bounds}
  
  We list three assumptions on $G_X$ and its ground state, $\ket{\zeta_X}$.

  \label{sec:mainresult}
   {\bf Assumption 1}:
  \begin{align}
  \nonumber
p_{\nu, \zeta_{X}}   \in \Theta(1) \; \forall X \; .
 \end{align}
 That is, there exists an $X$-independent state $\ket \nu$ that can be projected into $\ket {\zeta_{X}}$,
 with high probability,
 after a  measurement.
  
 {\bf Assumption 2}:
  \begin{align}
  \nonumber
p_{X, \zeta_{X}}  \in \Theta(1) \; \forall X \; .
 \end{align}
 That is,   $\ket {\zeta_{X}}$ can be projected into $\ket X$, with high probability,
 after a measurement. Assumptions 1 and 2 result in a probability of success $p_s \in \Theta(1)$
 when solving SEARCH with the generalized measurement-based method of Table I.
 
 Assumptions 1 and 2 may be combined into one as described in Appendix~\ref{appendix0}.
 Also, a generalization of Assumption 2 to any circuit $\cU$ is a requirement of
 having a ground state with large probability of being in the state output by the circuit
 after measurement. This property is
 desired for Hamiltonians 
 involved in the adiabatic simulation of quantum circuits.
 
 {\bf Assumption 3}:  For all $t \in \mathbb{R}$ and fixed $\epsilon$,  $0 \le \epsilon<1$, there exists a
 unitary operation ${ W_X}=(S.\tilde O_X)^r$,
  where $S$ is also a unitary operation that does not depend on $X$, 
  $ \tilde O_X = O_X \otimes \one$ is the oracle for SEARCH acting on the larger
  Hilbert space ${\cal H}'$, 
  $r \le   |c' t|^{\gamma}$, 
  and
  \begin{align}
  \nonumber
  \| e^{i G_X t} - W_X \| \le \epsilon \; .
  \end{align}
 $c'>0$ and $\gamma \ge 0$ are constants. 

Assumption 3 implies that the evolution operator determined by $G_X$ can be approximated,
 at precision $\epsilon$, by a sequence of unitary operations that uses the oracle  
order $|c' t|^\gamma$ times. For some specific $G_X$, such an approximation may follow from the results in Refs.~\cite{cleve_query_2009,berry_efficient_2007,wiebe_product_2010} on Hamiltonian simulation (see Sec.~\ref{sec:discussion}).

  {\bf Theorem.} If $G_X$ and $\ket{\zeta_X}$
  satisfy Assumptions 1, 2, and 3,   
  \begin{align}
  \nonumber
  \Delta^{\cU_X} \in \cO(1/L^{1/\gamma}) \; .
  \end{align}
  In addition, if $G_X$ satisfies a frustration-free property~\cite{bravyi_stoquastic_2009,somma_gap_2013},
   \begin{align}
   \nonumber
  \Delta^{\cU_X} \in \cO(1/L^{2/\gamma}) \; .
  \end{align}
  
  The definition of a frustration-free Hamiltonian is included in the proof.
  The second bound applies under an additional requirement on $G_X$. This requirement
  together with the constants for the upper bounds are also discussed in the proof. 
  
  We note that the gap in the second upper bound 
  may not be the ``relevant'' gap for the adiabatic simulation.
  In certain cases, for example, the adiabatic simulation 
  may not allow for transitions from the ground state to the first-excited 
  state due to symmetry reasons. Nevertheless, the first bound 
  still holds for the relevant gap in these cases.

  {\bf Proof.} 
Simulating the measurement in step ii of the generalized measurement-based method
 requires an evolution time  $T = c /\Delta^{\cU_X}$ [Eq.~\eqref{eq:GMBcost}].
 From Assumption 3, the evolution can be approximated by a quantum circuit that uses the oracle
 $r$ times, with $r \le (c'T)^\gamma$.
%
 The lower bound on the cost of solving SEARCH~\cite{bennett_searchbound_1997} implies
 \begin{align}
 \nonumber
  \left(\frac{c' c}{ \Delta^{\cU_X}} \right)^\gamma \ge r \ge \alpha  \sqrt N \ge \alpha 2L\; ,
 \end{align}
 where $\alpha>0$ is a constant because $p_s \in \cO(1)$.
 Then, $\Delta^{\cU_X} \le c' c/(2 \alpha L)^{1/\gamma}$.
 
  $G_X$ is a  frustration-free Hamiltonian if it is a sum of positive semidefinite
    terms  and the ground state  $\ket{\zeta_X}$ is a ground state of every term~\cite{bravyi_stoquastic_2009,somma_gap_2013,feiguin_renorm_2013}.
    In this case, it is possible to preprocess $G_X$ and build a Hamiltonian $\tilde G_{X}$ that has
    \begin{align}
    \label{eq:modHgroundstate}
   | \tilde \zeta_X \rangle = \ket{\zeta_X} \otimes \ket 0_{\rm a}
    \end{align}
   as (unique) eigenstate of eigenvalue zero, 
   where $\ket 0_{\rm a}$ denotes some simple, $X$-independent state of an ancillary system a.
   The corresponding spectral gap of $\tilde G_X$ for this state is $\tilde \Delta^{\cU_X} \ge \sqrt{ \Delta^{\cU_X}}$ -- see Ref.~\cite{somma_gap_2013} for details on spectral gap amplification.
      Then, SEARCH can be solved with probability $p_s \in \cO(1)$,
      using the generalized measurement-based method,
      by evolving with $\tilde G_X$ for time $T =c /\sqrt{ \Delta^{\cU_X}}$
     ~\footnote{
      To be rigorous, the state $\ket \nu$ has to be redefined as $\ket \nu \otimes \ket 0_{\rm a}$ for this case.}.
      If Assumption 3 also applies for approximating the evolution operator $e^{-i \tilde G_Xt}$, 
     then $\sqrt{ \Delta^{\cU_X}} \le c' c/(2 \alpha L)^{1/\gamma}$. This completes the proof.
  \vspace{0.3cm}
 
 {\bf Corollary.}  If $\gamma=1$, then $\Delta^{\cU_X} \in \cO(1/L)$. In addition,  if $G_X$ is frustration free
 as explained above, $\Delta^{\cU_X} \in \cO(1/L^2)$.
 
 \vspace{0.3cm}
 It is possible to achieve $\gamma \rightarrow 1$ for some $G_X$ (see Sec.~\ref{sec:discussion}).
 
 \vspace{0.3cm}
  {\bf Corollary.}  If the eigenvalues of $H^\cU$ do not depend on $\cU$,  the upper bounds
  on $\Delta^{\cU_X}$ are upper bounds on $\Delta^\cU$.

\section{Discussion: Validity of the Assumptions and  Implications}
\label{sec:discussion}
We  review the validity of the assumptions and  implications for some
constructions found in the literature.
The first is the standard construction in  Ref.~\cite{aharonov_adiabatic_2007}, also discussed in Sec.~\ref{sec:intro}.
In this case, we consider a modification of Grover's algorithm so that $\cU_X = \one^{L/4} (RO_X)^{L/4} \one^{L/4}$,
with $L \in \Theta(\sqrt N)$ and $\one$ the trivial (identity) operation.
Such a modification is unnecessary but it simplifies the analysis below. The state output by the modified circuit
is unchanged; the only change is in the Hamiltonians. As before, we let $G_X =H^{ \cU_X}$ be the Hamiltonian
associated with $\cU_X$ and $\ket{\zeta_X}= \ket{\psi^{\cU_X}}$ be its ground state [i.e., the history state of Eq.~\eqref{eq:historystate} with $\cU=\cU_X$]. For the modified circuit, the ground state has large overlap
with the $X$-independent state
\begin{align}
\nonumber
\ket \nu = \ket {\phi^0} \otimes \frac 1 {\sqrt{L/4+1}} \sum_{l=0}^{L/4} \ket l_{\rm c} \; .
\end{align}
Similarly, $\ket{\zeta_X}$ has large overlap with the state
\begin{align}
\nonumber
 \ket { X} \otimes \frac 1 {\sqrt{L/4}}   \sum_{l=3L/4}^{L} \ket l_{\rm c} \; ,
\end{align}
because $\ket X \approx  \ket{\phi^L_X} $ [see Eq.~\eqref{eq:halfgroverstate}].
These Eqs. imply $p_{\nu,\zeta_X} \approx 1/4$ and $p_{X,\zeta_X} \approx 1/4$,
so that Assumptions 1 and 2 are readily satisfied. 
%
%
%
%
%
%
%
To study Assumption 3, we write 
\begin{align}
\nonumber
G_{X}&= - O_X \otimes \sum_{l : U^l = O_X} [ \ket {l}\! \bra{l-1}_{\rm c} + \ket {l-1} \! \bra{l}_{\rm c}] +\ldots \\
\label{eq:Hrep}
&=   O_X \otimes P_{\rm c} +  H_{\rm s-c} \; .
\end{align}
 $P_{\rm c}$ is a Hamiltonian acting on the clock register that is a sum of commuting terms like
  $\ket {l}\! \bra{l-1}_{\rm c} + \ket {l-1} \! \bra{l}_{\rm c}$:  the oracles $O_X$ are interleaved with the 
  operations $R$
  in Grover's algorithm. Then, the eigenvalues of $P_{\rm c}$ are $\pm 1$ and
 $\| P_{\rm c} \| \le 1$, where $\| . \|$ is the operator norm.
$H_{\rm s-c}$ is a system-clock Hamiltonian that does not depend on $X$:
$H_{\rm s-c}$ is a sum of $H_{\rm input}$ and those terms in $H_{\rm Feynman}^{\cU_X}$
that do not depend on $O_X$. 
Using the results in Ref.~\cite{cleve_query_2009}, 
the operator $\exp\{iG_Xt\}$
can be well approximated using $ \cO(|t| \log|t|)$ oracles $O_X$ (see Appendix~\ref{appendixA}). 
  Thus, 
Assumption 3 is satisfied for the  construction of  Ref.~\cite{aharonov_adiabatic_2007} and $\gamma \rightarrow 1$ assymptotically.

To prove that $G_X$ is frustration free, we note that
\begin{align}
\label{eq:modHtransf}
G_X = W(\cU_X)^{\;} H^\one W(\cU_X)^{\dagger} \; , 
\end{align}
where $H^\one$ is the Hamiltonian of Eq.~\eqref{eq:standardH} for the trivial circuit and 
\begin{align}
\label{eq:modHtransf2}
W(\cU_X)^{\;} = \sum_{l=0}^L U^l \otimes \ketbra l_{\rm c} 
\end{align}
is a unitary operation. For the modified Grover's algorithm, $U^l \in \{ \one, R, O_X \}$.
It is simple to verify that $\ket{\psi^\one} \propto \ket{\phi^0}\sum_l \ket l_{\rm c}$, $h^{\one,l} \ket{\psi^\one} =0$, $h^{\one,l} \ge 0$, 
$H_{\rm input} \ket{\psi^\one}=0$, $H_{\rm input} \ge 0$. This implies that $H^\one$
is frustration free and so are $G_X$ and $H^\cU$ for any $\cU$. Then, there exists
\begin{align}
\nonumber
\tilde G_X = W(\cU_X)^{\;} \tilde H^\one W(\cU_X)^{\dagger} 
\end{align}
whose
ground state is $|\tilde\zeta_X \! \rangle=\ket{\zeta_X} \otimes \ket 0_{\rm a}$ and 
whose gap is $\sqrt{\Delta}$ \footnote{While the subspace of eigenvalue zero of $\tilde G_X$ is highly degenerate, the degeneracy is irrelevant
and can be easily removed by adding other $X$-independent terms~\cite{somma_gap_2013}.}. 
a is an ancilliary system of dimension $L+n$. The operators $h^{\cU,l}$
have eigenvalues $0,1$ and  $\sqrt{h^{\cU,l}}=h^{\cU,l}$.
Then, from the results in Ref.~\cite{somma_gap_2013}, Sec. IV,
we obtain
\begin{align}
\label{eq:modH1}
\tilde G_X = \tilde H^{\cU_X}_{\rm Feynman} + \tilde H_{\rm input} \; ,
\end{align}
with
\begin{align}
\nonumber
&\tilde H^{\cU_X}_{\rm Feynman} =  \sum_{l=1}^L h^{\cU_X,l} \otimes [\ket l \! \bra 0_{\rm a} + \ket 0 \! \bra l_{\rm a} ] \; , \\
\nonumber
&\tilde H_{\rm input} = \sum_{j=1}^n \ketbra- _j \otimes \ketbra 0_{\rm c} \otimes \\
\nonumber & \ \ \ \ \ \ \ \ \ \ \ \ \ \ \ \ \ \otimes [\ket {L+j} \! \bra 0_{\rm a} + \ket 0 \! \bra {L+j}_{\rm a} ] \; .
\end{align}
%
%
When $U^l=O_X$ in the modified Grover's algorithm,
\begin{align}
\nonumber
h^{\cU_X,l} = \frac 1 2 [\one( \otimes \ketbra l_{\rm c} + \ketbra{l-1}_{\rm c} ) + \\
\nonumber
+O_X \otimes \ket l \bra{l-1}_{\rm c} + \ket{l-1}\bra l_{\rm c}] \; .
\end{align}
Thus, another representation for $\tilde G_X$ is
\begin{align}
\label{eq:modHrep}
& \tilde G_X =  O_X \otimes \tilde P_{\rm c-a} + \tilde H_{\rm s-c-a} \; ,
\end{align}
with $\| \tilde P_{\rm c-a} \| \le 1$ because $\| \tilde H^\cU_{\rm Feynman} \| \le 1$
(see Appendix~\ref{appendix1}). 
The system-clock-ancilla Hamiltonian  $\tilde H_{\rm s-c -a}$ is independent of $X$.
Then, the evolution operator $e^{i\tilde G_X t}$ can be approximated from the results
in Ref.~\cite{cleve_query_2009} using the oracle $\cO(|t| \log|t|)$ times and the
gadget in Appendix~\ref{appendixA}. 
It follows that $\gamma \rightarrow 1$ asymptotically for this case as well, and the gap satisfies $\Delta^{\cU_X} \in \tilde \cO(1/L^2)$. (The $\tilde \cO$ notation accounts for the additional  logarithmic factor.)
This upper bound is also valid for any $\Delta^\cU$, because the eigenvalues of $H^\cU$ do not depend
on $\cU$ [Eq.~\eqref{eq:modHtransf}].
Our result is compatible with the lower bound on $\Delta^\cU$ obtained in Ref.~\cite{aharonov_adiabatic_2007} (see Sec.~\ref{sec:intro}).
It proves that our technique to establish limits in the gap is effective. Nevertheless, as we show below, 
our technique is powerful when analyzing the gaps of Hamiltonians that are simple modifications
to the $G_X$ above, where obtaining the spectrum  directly
can be challenging. We note again that, since the local Hamiltonian constructed in Ref.~\cite{aharonov_adiabatic_2007}
has a smaller gap than that of $H^\cU$ or $G_X$, the bound on the gap of $G_X$ translates into a bound
on the gap of the local Hamiltonian.

%
%

We  use the previous analysis  
to show a more general result.
Consider a general Hamiltonian $H^\cU=W(\cU) H^\one W(\cU)^\dagger$ for the adiabatic
simulation of a quantum circuit, which uses a clock register,
 and whose ground state is of the form
 \begin{align}
 \nonumber
 \ket{\psi^\cU}& = W(\cU) \ket{\psi^\one} \\
  \label{eq:generalhistory}
 & = \sum_{l=0}^L \alpha^l \ket{\phi^l} \otimes \ket l_{\rm c} \; ,
 \end{align}
and $\ket{\psi^\one} = \ket{\phi^0} \otimes \sum_l \ket l_{\rm c}$.
With no loss of generality, we can assume that
there exists $l_0$ such that
\begin{align}
\label{eq:amplitudecondition}
\sum_{l=l_0}^L |\alpha^l|^2 \in \Theta(1) \; .
\end{align}
If this condition is not satisfied, we can always apply
an operation that permutes the clock states or we can add trivial
operations to the circuit so that Eq.~\eqref{eq:amplitudecondition}
is satisfied (the spectrum of $H^\cU$ is unchanged). We let $l_0$ be the largest $l$ to satisfy Eq.~\eqref{eq:amplitudecondition}.
Then, we consider a modification of Grover's algorithm so that
\begin{align}
\nonumber
\cU_X = \ketbra 0_{\rm b} \otimes \one + \ketbra 1_{\rm b} \otimes \left(\one^{l_0} .(RO_X)^{L-l_0}\right) \; ,
\end{align}
where b is an ancillary qubit (see Appendix~\ref{appendix0}). $L \in \Theta(\sqrt N)$. Basically, the modified Grover's algorithm
acts trivially, if the state of an ancillary qubit is $\ket 0_{\rm b}$, or implements the original
Grover's algorithm, if the state of the ancilla is $\ket 1_{\rm b}$. The initial state is $\ket+_{\rm b} \otimes \ket{\phi^0}$,
and $\ket{\phi^0}$ is the equal superposition state as required in Grover's algorithm. 
Assumptions 1 and 2 then follow from Eq.~\eqref{eq:amplitudecondition}, for those ground states that
can be described by Eq.~\eqref{eq:generalhistory}.
Additionally, if the Hamiltonian associated with $\cU_X$ can be represented as in Eq.~\eqref{eq:Hrep},
with $\| P_{\rm c} \| \le 1$,
the evolution under $G_X$ can be well approximated using $\cO(t \log t)$ oracles
and the upper bound on $\Delta^{\cU_X}$ is of $\tilde \cO(1/L)$. 
Such Hamiltonians include those $H'^{\cU}$
arising from modified Feynman Hamiltonians, where $H'^\cU_{\rm Feynman} = \sum_l \beta^l
h^{\cU,l}$, $|\beta^l | \le 1$, and those Hamiltonians that have an additional term
\begin{align}
\nonumber
H_{\rm pointer} = \sum_l E^l . \one \otimes \ketbra l_{\rm c} \; ,
\end{align}
that acts solely in the clock space.

For those $H'^\cU$, the spectrum is independent of $\cU$ (i.e., $\Delta^{\cU_X}=\Delta^\cU$) and, in particular,
\begin{align}
\nonumber
G_X = W^{\!}(\cU_X) H'^\one W(\cU_X)^\dagger \; 
\end{align}
[see Eqs.~\eqref{eq:modHtransf} and~\eqref{eq:modHtransf2}]].
The unitaries $U^l$ involved in the definition of $W(\cU_X)$
are  $U^l \in \{ \one, \ketbra 0_{\rm b} \otimes \one + \ketbra 1_{\rm b} \otimes O_X,
 \ketbra 0_{\rm b} \otimes \one + \ketbra 1_{\rm b} \otimes R \}$, for the current 
 $\cU_X$. $H'^\one$ acts trivially in the system and has tridiagonal form
 in the basis $\{\ket 0 _{\rm c} , \ldots, \ket L_{\rm c} \}$. Then,
 with no loss of generality, we can assume that $H^\one$ is frustration free \footnote{
 The frustration free property  can be obtained by adding, for example, a constant to $H^\one$ so that
 its lowest eigenvalue is zero.}. It follows that $G_X$ is also frustration free and we can build
 \begin{align}
 \nonumber
\tilde G_X = W^{\;}(\cU_X) \tilde H^\one W^\dagger(\cU_X) \; ,
\end{align}
 by using the results of Ref.~\cite{somma_gap_2013}. Because $\tilde H^\one$
 is also tridiagonal in the basis $\{\ket 0 _{\rm c} , \ldots, \ket L_{\rm c} \}$,
 $\tilde G_X$  admits a representation of the form of Eq.~\eqref{eq:modHrep}
 in this case, with $\|\tilde P_{\rm c-a}\| \le 1$.
%
%
%
Then, the oracle cost of simulating $\tilde G_X$ for time $t$ is also $\cO(t \log t)$.
 This implies that, for modified Feynman Hamiltonians, the second bound on the gap applies with $\gamma \rightarrow 1$,
 and $\Delta^\cU \in \tilde \cO(1/L^2)$.

 A few remarks are in order. First, we note that the above result contradicts a statement in Ref.~\cite{lloyd_adiabatic_2008}
claiming that the gap can be amplified to order $1/L$ by including a term
of the form $H_{\rm pointer}$.
 Second, that an upper bound on $\Delta^\cU$ of order $1/L^{2/\gamma}$ is obtained
 when the Hamiltonian satisfies the frustration free property, 
 it does not contradict
 that, for some Hamiltonians, the ``relevant'' gap in certain subspace (e.g., the 
 translationally invariant subspace) may be larger.  Nevertheless, such a relevant gap
 should be limited by the bound on $\Delta^\cU$ obtained without assuming the property in the frustration
 (i.e., $1/L^{1/\gamma}$ in this case).  
 A third remark concerns the applicability of our results to
 those constructions in which the Hamiltonians are associated with one-dimensional quantum systems,
 such as the one in Ref.~\cite{aharonov_line_2009}. These constructions would require
 ``breaking'' the oracle $O_X$ into local, two-qubit pieces. While Assumptions 1 and 2 are easy to verify, 
 a new version of Assumption 3 is required for this case. Such a version may be possible   even if
 the oracle is now a composition of two-qubit local operations, because the evolution operator with the one-dimensional
 Hamiltonian may ``reconstruct'' a full oracle after certain unit of evolution time. However, we do not have
 any rigorous result for this case and finding other suitable versions of Assumption 3 is work in progress. 
 Finally, Assumptions 1 and 2 do not apply to the construction in A. Mizel, e-print: arXiv:1002.0846 (2010).

 \section{ Acknowledgements}
 We  thank S. Boixo, R. Blume-Kohout, D. Gossett, A. Landahl, and D. Nagaj for insightful discussions. 
We  acknowledge support from  the Laboratory Directed Research and Development Program at 
Sandia National Laboratories.
 Sandia National Laboratories is a multi-program laboratory managed and operated by Sandia Corporation, a wholly owned subsidiary of Lockheed Martin Corporation, for the U.S. Department of Energy's National Nuclear Security Administration under contract DE-AC04-94AL85000.

\hspace{1cm}
 
 \begin{appendix}
 
  \section{More on Assumptions 1 and 2}
 \label{appendix0}
 In general, Assumption 2 is mostly an statement about the ground state of the Hamiltonian $H^\cU$
 that simulates a quantum circuit, $\ket{\psi^\cU}$. Ideally, such state has large probability of being in the state output
 by the circuit, $\ket{\phi^L}$; that is,
 \begin{align}
 \nonumber
 \Pr({\phi^L|\psi^\cU})= \Tr [\bra{\phi^L}\psi^\cU \rangle \langle \psi^\cU \ket{\phi^L} ]  \in \Theta(1) \; .
 \end{align}
 We can then consider a modified quantum circuit that uses an additional ancilla b
 prepared in $\ket +_{\rm b}$ so that it applies the unitary $\cU$ (original circuit)
 controlled on the state $\ket 1$
 of the ancilla, or does nothing otherwise.
 If we denote the modified circuit by $\bar \cU$, the output state is
 \begin{align}
 \nonumber
 \ket{\bar \phi^L}&= \bar \cU \left ( \ket + \otimes \ket{\phi^0} \right) \\
 \nonumber
 & =\frac 1 {\sqrt 2} [\ket 0_{\rm b} \otimes \ket{\phi^0} + \ket 1_{\rm b} \otimes \ket { \phi^L}]\; .
 \end{align}
 In this way, if the ground state of $H^\cU$ is a superposition of system-clock states of the form
 $\ket {\phi^l} \otimes \ket l_{\rm c}$, the ground state of $ H^{\bar \cU}$ will be a superposition 
 of states of the form $\ket {\bar \phi^l} \otimes \ket l_{\rm c}$, with $\ket{\bar \phi^l} = \bar U^l \cdots \bar U^0 (\ket +_{\rm b} \otimes \ket {\phi^0})$. When $\cU=\cU_X$
 corresponds to Grover's algorithm, if $|\psi^{\bar \cU}\rangle$
 has large probability of being in $| \bar \phi^L\rangle$ after measurement, 
 then it has large probability of being in both, $\ket{\phi^0}$ and $\ket{\phi^L}$,
 after respective measurements. Since $\ket{\phi^0}$ is independent of $X$,
 $| \bar \phi^L \rangle$ satisfies Assumption 1 and 2 simultaneously.
 Thus, 
 in Grover's algorithm, Assumptions 1 and 2 can be combined into a single one
 for Hamiltonians whose ground states are superpositions of $\ket {\phi^l} \otimes \ket l_{\rm c}$.
 The gap bounds will apply to $H^{\bar \cU_X}$ in this case.
 
 \section{Oracle simulation of the Feynman Hamiltonian associated with Grover's algorithm}
 \label{appendixA}
 Following Ref.~\cite{cleve_query_2009},
 the first step is to use the Trotter-Suzuki approximation that,
 in the case of
  the evolution 
 under $G_X=O_X \otimes P_{\rm c} + H_{\rm s-c}$,
it  yields terms of the form  
 \begin{align}
 \label{eq:coupledoracle}
 e^{-i s O_X \otimes P_{\rm c}}
 \end{align}
 for some small $s \in \mathbb{R}$. 
 The goal in this section is to present gadget that implements Eq.~\eqref{eq:coupledoracle}
 (i.e., a fractional oracle) using $O_X$. Then, the problem is reduced to the one analyzed in
 Ref.~\cite{cleve_query_2009}, for which the oracle cost is known.
 
 First, we note that there exists a unitary 
 operation $V_{\rm c}$ such that
  \begin{align}
 \label{eq:coupledoracle2}
V_{\rm c}^{\;} e^{-i s O_X \otimes P_{\rm c}} V_{\rm c}^\dagger = e^{-i s O_X \otimes D_{\rm c}} \; ,
 \end{align}
 where $D_{\rm c}$ is a diagonal operator acting on the clock register, i.e.,
 \begin{align}
 \nonumber
 D_{\rm c} = \sum_k \lambda_k \ketbra  k_{\rm c} \; ,
 \end{align}
 and $|\lambda_k| \le 1$ because $\|P_{\rm c}\| \le 1$.
 $V_{\rm c}$ commutes with $O_X$ and it does not depend on $X$.
  The  ``gadget'' of Fig.~\ref{fig:exact} uses this observation to implement the operation of the {\em rhs} of  
   Eq.~\eqref{eq:coupledoracle2}. Then, the desired operator of Eq.~\eqref{eq:coupledoracle} can be implemented   by conjugating the circuit of Fig.~\ref{fig:exact} with $V_{\rm c}$. This has to be compared with Fig. 3 of Ref.~\cite{cleve_query_2009}.
   \begin{figure}[htb]
\vspace{-.4cm}
\begin{equation*}
\Qcircuit @C=.7em @R=.4em{ & \\
&  & \lstick{\ket{k}_{\rm c}} &  \ctrl{1} & \qw & \ctrl{1} & \qw&  \rstick{\! \!  \!  \ket{k}_{\rm c}}  & \push{\rule{0em}{2em}} \\
&  & \lstick{\ket{0}_{\rm b}} & \gate {R_1} & \ctrl{1} & \gate{R_2}  & \meter \\
& \qw& \qw & \qw& \multigate{3}{O_X} & \qw   & \qw \\
& \qw& \qw & \qw& \ghost{O_X} & \qw  & \qw  \\
& \qw& \qw & \qw& \ghost{O_X} & \qw   & \qw  \\
& \qw& \qw & \qw& \ghost{O_X} & \qw  & \qw  \\
}
\end{equation*}
\caption{Simulation of  $\exp\{-isO_X \otimes D_{\rm c}\}$ [Eq.~\eqref{eq:coupledoracle2}].
b is an ancilla qubit. The controlled operations are: $R_1 \ket0_{\rm b} \propto \sqrt{\cos (s \lambda_k/2)}
\ket0_{\rm b} - i \sqrt{\sin (s \lambda_k/2)} \ket 1_{\rm b}$ and
$R_2 \ket {0 (1)}_{\rm b} \propto \sqrt{\cos (s \lambda_k/2)}
\ket0_{\rm b} -(+)  \sqrt{\sin (s \lambda_k/2)} \ket 1_{\rm b}$ (see Fig. 3 in Ref.~\cite{cleve_query_2009}).
The ancilla is measured at the end and the simulation of $e^{-is \lambda_k O_X \otimes \ketbra k_{\rm c} }$
succeeds if the outcome is  $\ket 0_{\rm b}$. The oracle is controlled in the state $\ket 1_{\rm b}$.}
\label{fig:exact}
\end{figure}

If we use the simulation of Fig.~\ref{fig:exact} in the scheme shown in Fig.4 of Ref.~\cite{cleve_query_2009}, 
the total number of oracles needed for approximating the evolution operator $e^{-i G_X t}$
is of order $\cO(|t| \log|t|)$. This requires implementing other simulation ``tricks'' to reduce the oracle cost, such as 
reducing the Hamming weight of the state of the ancillas for each simulation of 
$e^{-i s O_X \otimes D_{\rm c}}$,  coming from the Trotter-Suzuki approximation (see Ref.~\cite{cleve_query_2009}
for more details).

 
   \section{The modified Hamiltonians $\tilde G_X$}
 \label{appendix1}
 The first modified Hamiltonian we analyze is the one in Eq.~\eqref{eq:modH1}
 for Grover's algorithm, and write $\tilde G_X = \tilde H^{\cU_X}$.
 Then,
 \begin{align}
\nonumber
& \tilde G_X =  \\
\nonumber
& =- O_X \otimes \sum_{l : U^l=O_X} [ \ket {l}\! \bra{l-1}_{\rm c} + \ket {l-1} \! \bra{l}_{\rm c}] \otimes \\
\nonumber
& \otimes[ \ket{l}\bra 0_{\rm a} + \ket 0 \bra{l}_{\rm a}]+ \ldots \\
\nonumber
&= O_X \otimes \tilde P_{\rm c-a} + \tilde H_{\rm s-c-a} \; .
\end{align}
$H_{\rm s-c-a} $ is a Hamiltonian that 
contains terms of the system, clock, and ancilla a
not included in the first term.
It does not contain any term that depends on $O_X$, i.e., it contains only those with $R$
(and $\one$ for the modified algorithm).
Because the set $\{ l : U^l=O_X \}$ involves only odd or even values of $l$  (i.e., 
$R$ and $O_X$ alternate in Grover's algorithm), the operator $P_{\rm c-a}$
is a sum of commuting terms, each of the form
\begin{align}
\nonumber
-   [ \ket {l}\! \bra{l-1}_{\rm c} + \ket {l-1} \! \bra{l}_{\rm c}] \otimes [ \ket{l}\bra 0_{\rm a} + \ket 0 \bra{l}_{\rm a}] \; .
\end{align}
The eigenvalues of each of these terms are $\pm 1$, implying that $\| P_{\rm c-a} \| =1$.
 
%
%
%

 \end{appendix}
  
  \newpage
 

\end{document}